\documentclass[a4paper]{article}
\usepackage{amsmath,graphicx}
\usepackage{amsfonts}
\usepackage{amssymb}
\usepackage{geometry}

\usepackage{caption}
\usepackage{subcaption}

\date{}

\title{Joint Device-Edge Inference over Wireless Links with Pruning}

\author{Mikolaj Jankowski, Deniz G{\"u}nd{\"u}z, Krystian Mikolajczyk\\
Imperial College London\\
{\tt\small \{mikolaj.jankowski17, d.gunduz, k.mikolajczyk\}@imperial.ac.uk}}

\begin{document}

\maketitle

\begin{abstract}
We propose a joint feature compression and transmission scheme for efficient inference at the wireless network edge. Our goal is to enable efficient and reliable inference at the edge server assuming limited computational resources at the edge device. Previous work focused mainly on feature compression, ignoring the computational cost of channel coding. We incorporate the recently proposed deep joint source-channel coding (DeepJSCC) scheme, and combine it with novel filter pruning strategies aimed at reducing the redundant complexity from neural networks. We evaluate our approach on a classification task, and show improved results in both end-to-end reliability and workload reduction at the edge device. This is the first work that combines DeepJSCC with network pruning, and applies it to image classification over the wireless edge.
\end{abstract}


\section{Introduction}

Number of Internet of things (IoT) devices has reached 22 billion at the end of 2018, and is expected to grow up to 75 billion by the end of 2025 \cite{IHS_report_iot}. 
Currently, most of these devices act as wireless sensors that collect data and offload it to a cloud or edge server for processing. This creates a major bottleneck in many emerging IoT applications as communication consumes significant energy and introduces errors and latency.

In this work, we consider deep neural network (DNN) based inference at an edge device \cite{Gunduz_JSAC}. Due to limited computational power and memory, IoT devices typically cannot perform all the computations required by a complex DNN architecture. For example, a single forward pass of the ResNet-152 \cite{resnet} architecture requires $11\times 10^9$ floating-point operations (FLOPs) for a $224\times 224$ input image. This would take few minutes on a simple IoT device, which is usually limited to a few MFLOPs per second. We assume that an edge server is available to help the device to perform the inference task. In most current implementations, the IoT device offloads all its data to the edge server, where a DNN of any complexity can be deployed. Note, however, that parts of the data that the device is sending  may not be useful for the underlying task. An alternative approach is to preprocess the data on the edge device, within the available computational limits, and transmit only the resulting features  to the edge server. We therefore encounter two main challenges: minimizing the number of computations that have to be done locally, and designing a robust communication scheme within the limited available transmission power and bandwidth.


\begin{figure}[h]
\begin{center}
\includegraphics[width=0.99\textwidth]{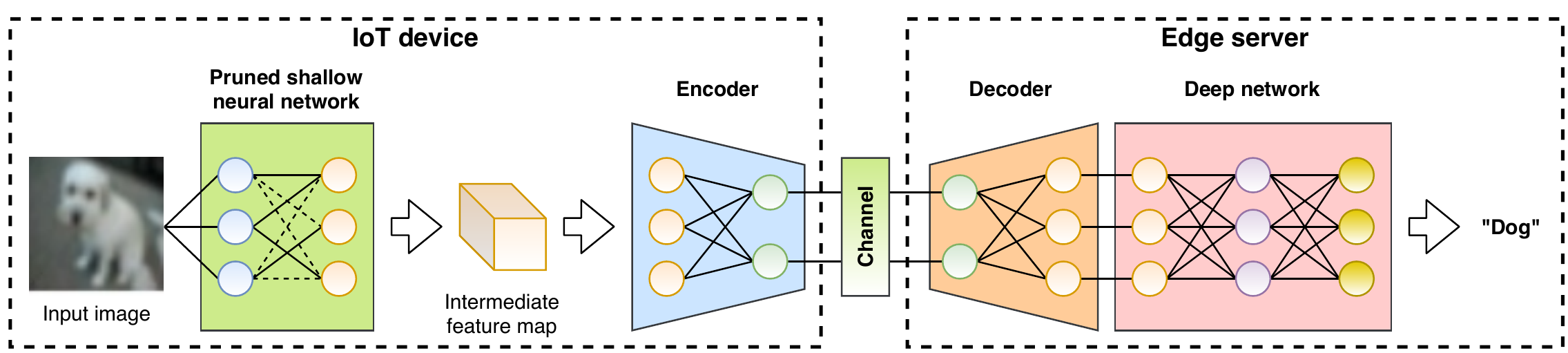}
\end{center}
\caption{An overview of the proposed system. The baseline neural network is split between an IoT device and an edge server. Given input data (e.g., an image), pruned sub-network performs the first part of the forward pass to generate an intermediate feature map, which is then compressed by DeepJSCC encoder and sent through a wireless link. At the receiver side, first the compressed feature map is reconstructed, and the remaining part of the forward pass is completed to obtain the final prediction.}
\label{fig:system}
\end{figure}

To address the first challenge, DNN architectures that operate within the low-complexity constraints of mobile devices are proposed in \cite{mobilenet}. These may still need hundreds of millions FLOPs to perform a single forward-pass, which may be unacceptable for certain IoT devices. Some recent works \cite{dnn_decoupling, bottlenet, bn_plus_plus} suggest splitting DNNs into two parts, where only the first few layers are implemented on the device within its computational constraints, while the remaining layers are deployed on the edge server. However, this approach requires reliable transmission of the intermediate feature vectors to the edge server. To reduce the communication requirements, a typical approach is to quantize and/or compress the feature vectors before transmitting over the channel \cite{dnn_decoupling, bottlenet}.  These methods consider the amount of information (e.g., the number of bits) that must be conveyed to the edge server, but ignore the energy and latency cost of communications, and potential errors that may be introduced. Moreover, reliable transmission of the feature vectors requires an accurate estimate of the channel state at the edge device, and separate compression and channel coding is known to be suboptimal under strict delay constraints. Recently, a DNN-based DeepJSCC scheme has been shown to provide improved performance and robustness in wireless image transmission \cite{jscc_image_dnn, david_spawc, jscc_feedback}. DeepJSCC scheme has been applied in distributed inference scenarios as well \cite{jscc_features, bn_plus_plus, jankowski2019deep}, but they require a significant number of on-device computations to run a forward pass of the underlying DNN.

In this work, we propose to reduce the on-device computational load by incorporating a pruning step in the network training. Network pruning \cite{optimal_brain_damage} aims at reducing the computational redundancy within DNNs by efficiently removing certain neurons, convolutional filters, or entire layers based on a saliency measure or a regularization term. Therefore, given a certain channel condition and a computational constraint at the IoT device, our goal is to find the optimal DNN splitting point and the pruning parameters to ensure the best possible accuracy as well as efficiency in a given scenario. We consider image classification as the target application. Network splitting together with pruning for edge devices has recently been studied in \cite{iot_pruning}, but similarly to other works, they ignore the errors that may be introduced over the channel. In contrast, our work is the first to study a joint device-edge inference architecture combined with pruning taking into account the noisy wireless channel. Our contributions can be summarized as follows:

\begin{itemize}
    \item We propose a DNN training procedure for joint device-edge inference systems under extreme power and latency constraints by combining novel pruning and splitting techniques with end-to-end feature transmission.
    \item Inspired by the DeepJSCC architecture \cite{jscc_image_dnn}, we propose an autoencoder-based network for intermediate feature map transmission to allow bandwidth reduction.
    \item We present extensive evaluations of the proposed approach at various DNN splitting points, channel SNRs, and bandwidth and computational power constraints, and show its efficacy in a wide range of settings. 
\end{itemize}

\section{Methods}

We propose a 4-step training strategy for combining partial network pruning with an end-to-end autoencoder architecture for transmitting intermediate feature maps of an arbitrary hidden layer of a DNN (Fig. \ref{fig:system}). Such an approach allows for a reduction in the computations carried out at the IoT device, while taking into account the effect of channel noise on the performance (within the specified bandwidth constraint). Most popular convolutional neural networks (CNNs), such as VGG \cite{vgg} or ResNet \cite{resnet} perform spatial dimensionality reduction of intermediate feature maps by applying pooling operations or convolutional filters with stride greater than 1. Nevertheless, as spatial dimension is being reduced, number of channels is usually expanded to extract the most significant features from the input image. Therefore, in the first few layers of such networks, the total dimension of the feature map increases up to a certain point, after which it starts to decrease due to further downsampling. As a consequence, there exists a hidden layer within such a network, where the size of the intermediate feature map is lower than that of the input; which acts as data compression for the underlying task. Transmitting such a feature map instead of the original input can help to reduce the bandwidth, but will impose significant computational resources on the device as it will have to preprocess the image through many layers of the network. This defines a trade-off between on-device computation and communication bandwidth, which we aim to optimize.

\subsection{Channel model}

We consider an additive white Gaussian noise (AWGN) channel, but any differentiable channel model can be employed instead. Specifically, given a channel input vector $\mathbf{x} \in \mathbb{R}^{B}$, where $B$ represents the available channel bandwidth, the channel output $\mathbf{y} \in \mathbb{R}^{B}$ is given by $\mathbf{y} = \mathbf{x} + \mathbf{z}$, where $\mathbf{z}$ is an independent and identically distributed (i.i.d.) noise vector with elements $z\sim \mathcal{N}\left(0, \sigma^2\right)$. An average power constraint of  $P=1$ is imposed on the channel input vector, i.e., $\frac{1}{B}  \sum_{i=1}^B x_{i}^2 \leq P$. We evaluate the accuracy for different channel signal-to-noise ratios (SNRs) given by $\frac{P}{\sigma^2}$. To compare our JSCC approach to digital methods we use standard Shannon capacity formula given by $C = \frac{1}{2}\log_2\left(1 + \frac{P}{\sigma^2}\right)$.

\subsection{Classification baseline}

Our framework is flexible, and can be easily adapted to any system that incorporates DNNs. We focus on image classification task as it is the most frequent approach to automatically analyse image content and generate its metadata. Given an image and a finite set of possible classes, the classification task aims at assigning the correct class label to the image. We experiment with VGG16 \cite{vgg} with batch normalization (BN) added after each convolutional layer as it is one of the most popular networks employed for image classification. The network consists of 13 convolutional layers with stride 1 divided into 5 blocks, where each block is followed by a pooling operation. We consider each of the pooling operations as a potential network splitting point as it provides feature compression by construction, and does not affect the accuracy. After the last pooling layer we also employ a fully-connected classifier consisting of three fully-connected layers, where the first two have the output size of 512 and the last one maps 512-dimensional vector to 100-dimensional class predictions.

\subsection{Autoencoder architecture}

\begin{figure}[h]
    \centering
    \begin{subfigure}[b]{0.30\linewidth}
        \centering
        \includegraphics[width=\textwidth]{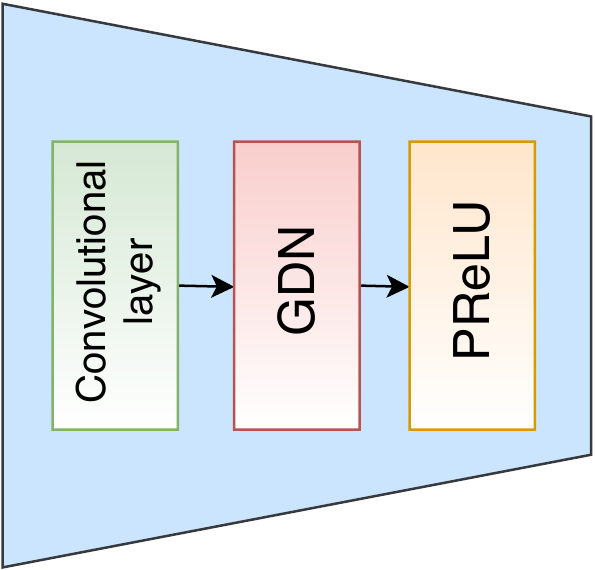}
        \caption{Encoder}
        \label{fig:encoder}
    \end{subfigure}
    \hspace{1pt}
    \begin{subfigure}[b]{0.66\linewidth}
        \centering
        \includegraphics[width=\textwidth]{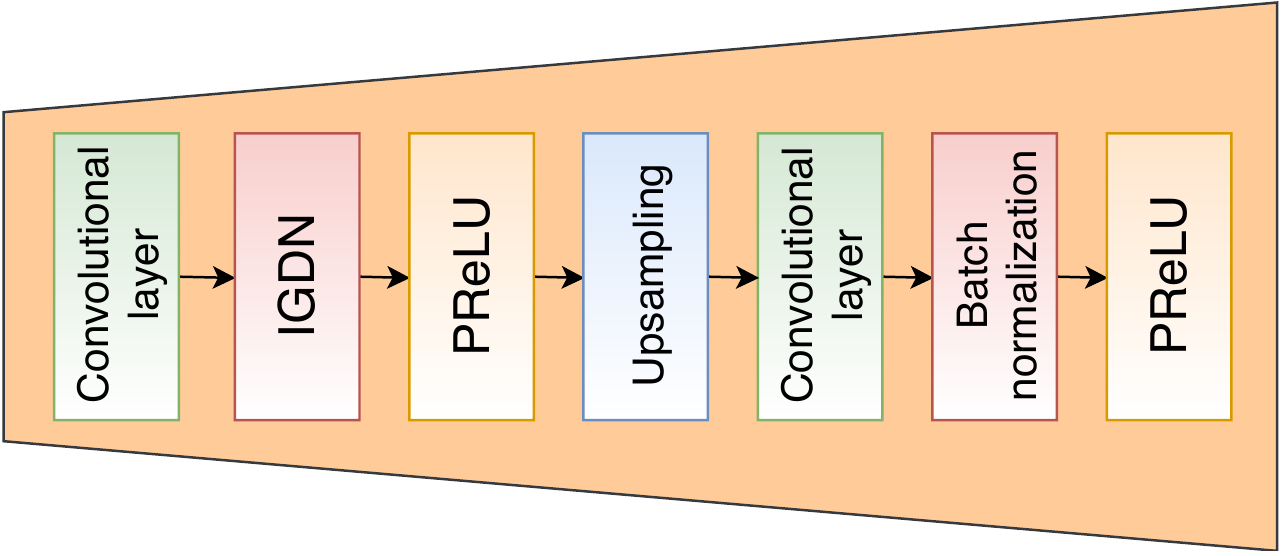}
        \caption{Decoder}
        \label{fig:decoder}
    \end{subfigure}
       
    \caption{Proposed encoder and decoder architecture for feature transmission. At the encoder, dimensionality reduction is performed by the convolutional layer. Shallow structure of the encoder reduces the computational load on the power-constrained device.}
    \label{fig:autoencoder}
\end{figure}

In this work, we explicitly model and evaluate the impact of the noisy wireless channel on the performance. Therefore, we design the communication scheme in conjunction with the DNN architecture employed for the underlying classification task. As opposed to most of the literature on device-edge co-inference, we do not employ digital codes to transmit the feature maps, which are known to be suboptimal in finite blocklengths. Instead, we employ the autoencoder architecture shown in Fig. \ref{fig:autoencoder}. Its asymmetrical structure is designed to reduce on-device computations; therefore, the encoder's architecture (Fig. \ref{fig:encoder}) consists of a single convolutional layer of stride $2\times 2$ and $3\times 3$ kernels, which perform both spatial and channel-wise compression of the feature map in a single step. The convolutional layer is followed by the generalized divisive normalization (GDN) layer \cite{gdn}, which is commonly used in most successful deep compression schemes such as \cite{DBLP:journals/corr/BalleLS16a} as a replacement of BN. GDN operation is defined as:
\begin{equation}
    u_i^{(k+1)}(m, n) = \frac{w_i^{(k)}(m, n)}{\left( \beta_{k, i} + \sum_j\gamma_{k, ij}\left(w_j^{(k)}(m, n)\right)^2\right)^{\frac{1}{2}}},
\end{equation}
where $u_i^{(k)}(m, n)$ denotes the $i$th output channel of the $k$th stage of the encoder at the spatial location $(m, n)$ and $w_i^{(k)}(m, n)$ denotes the corresponding input value. The approximate inverse operation, called IGDN is given by:
\begin{equation}
    \footnotesize{
    \hat{w}_i^{(k+1)}(m, n) = \hat{u}_i^{(k)}(m, n)\left( \hat{\beta}_{k, i} + \sum_j\hat{\gamma}_{k, ij}\left(\hat{u}_j^{(k)}(m, n)\right)^2\right)^{\frac{1}{2}},
    }
\end{equation}
where $\hat{w}$ and $\hat{u}$ are the output and input of IGDN, respectively. Finally, we employ parametric rectified linear unit (PReLU) \cite{prelu} as an activation function to further increase the learning capacity of our model. The output of the encoder network is directly transmitted over the channel (after normalization - to meet the power constraint).

At the decoder (Fig. \ref{fig:decoder}) we first perform a single convolution with stride $1\times 1$ and $3\times 3$ kernel size  on the compressed feature map.
This is followed by the IGDN operation, PReLU activation, and upsampling to restore the original spatial dimension of the intermediate feature map. Finally, another convolutional layer with the same stride and kernel size is applied to increase the feature map's depth to its original value, followed by BN and PReLU. Note that the number of channels effectively controls the size of the transmitted vector as our encoder always reduces the spatial dimensionality by a factor of 4. The only exception is the last block of VGG16 network (after pooling 5), where the feature map of size $1\times 1\times 512$ cannot be downsampled so we
only control the number of channels.

\subsection{Training strategy}
\label{subs:training_strategy}

Our training strategy consists of 4 steps. Firstly, we pretrain the VGG16 network with cross-entropy loss for 60 epochs, using SGD \cite{sgd} optimizer with a learning rate of $0.01$, momentum of $0.9$, and $L_2$ penalty on network parameters weighted by $5\cdot 10^{-4}$ to avoid overfitting. We reduce the learning rate by a factor of $0.1$ after 20th and 40th epochs.

Next, we select the splitting point after one of the pooling layers of the network and employ network pruning. We use the pruning algorithm in \cite{pruning}, which uses Taylor expansion to approximate the change in the loss function induced by pruning. In principle, the algorithm evaluates the importance of each convolutional filter up to the splitting point, and removes the least significant ones. In our setup, the algorithm removes 512 convolutional filters at a single pruning iteration, followed by additional 10 training epochs with a learning rate of $0.0001$ to recover the accuracy lost by the filter removal.

Afterwards, we run a forward pass through the pruned network with each image in the training set to extract all the possible feature maps at the splitting point. We use the feature maps as a new training set for our autoencoder, which we pretrain for 40 epochs with a learning rate of $0.1$, momentum of $0.9$, and $L_2$ penalty weighted by $5\cdot 10^{-4}$. We use $L_1$-loss to recover the feature maps as close to their original versions as possible. This step is crucial to speed-up the convergence of the end-to-end training; since the feature maps are low-dimensional and autoencoder architecture is very simple, its execution is relatively fast. Starting from this step, we incorporate an AWGN channel model between the encoder and decoder to gain robustness against channel noise.

In the last step, we perform end-to-end training of the entire network. Specifically, we combine both parts of the VGG16 network and place pretrained autoencoder at the splitting point. Similarly to the first step, we train the network with cross-entropy loss, SGD optimizer with learning rate of $0.0001$ and the other parameters unchanged.

\section{Results}

In this section we evaluate the performance of the proposed approach and compare it with other schemes in the literature. 

\subsection{Experimental setup}

In order to evaluate the accuracy of the proposed method, we employ popular CIFAR100 dataset, which consists of 60000 RGB images divided into 100 different classes (e.g., bicycle, fox, oranges, etc.) \cite{cifar}. Each class is represented evenly by 600 images of size $32\times 32$ pixels, 500 for training and 100 for testing. During training, we first perform common data augmentation steps, namely we apply 4 pixel zero-padding at each side of an image and randomly crop $32 \times 32$ pixel tiles. Moreover, we randomly flip images horizontally with a probability of $50\%$ and normalize them to have zero mean and unit variance. After such preprocessing, we perform multiple training runs of the proposed system, according to the strategy in Section \ref{subs:training_strategy} for different values of channel SNR, pruning ratios, network splits, and channel bandwidths, and evaluate the corresponding classification accuracy and required number of computations. In order to calculate the computational complexity of our approach, we count the number of FLOPs necessary to perform a single forward pass of the layers executed at the edge device (pruned shallow sub-network and the encoder).

\subsection{Channel bandwidth and on-device computation}

\begin{figure}[h]
\begin{center}
\includegraphics[width=0.8\linewidth]{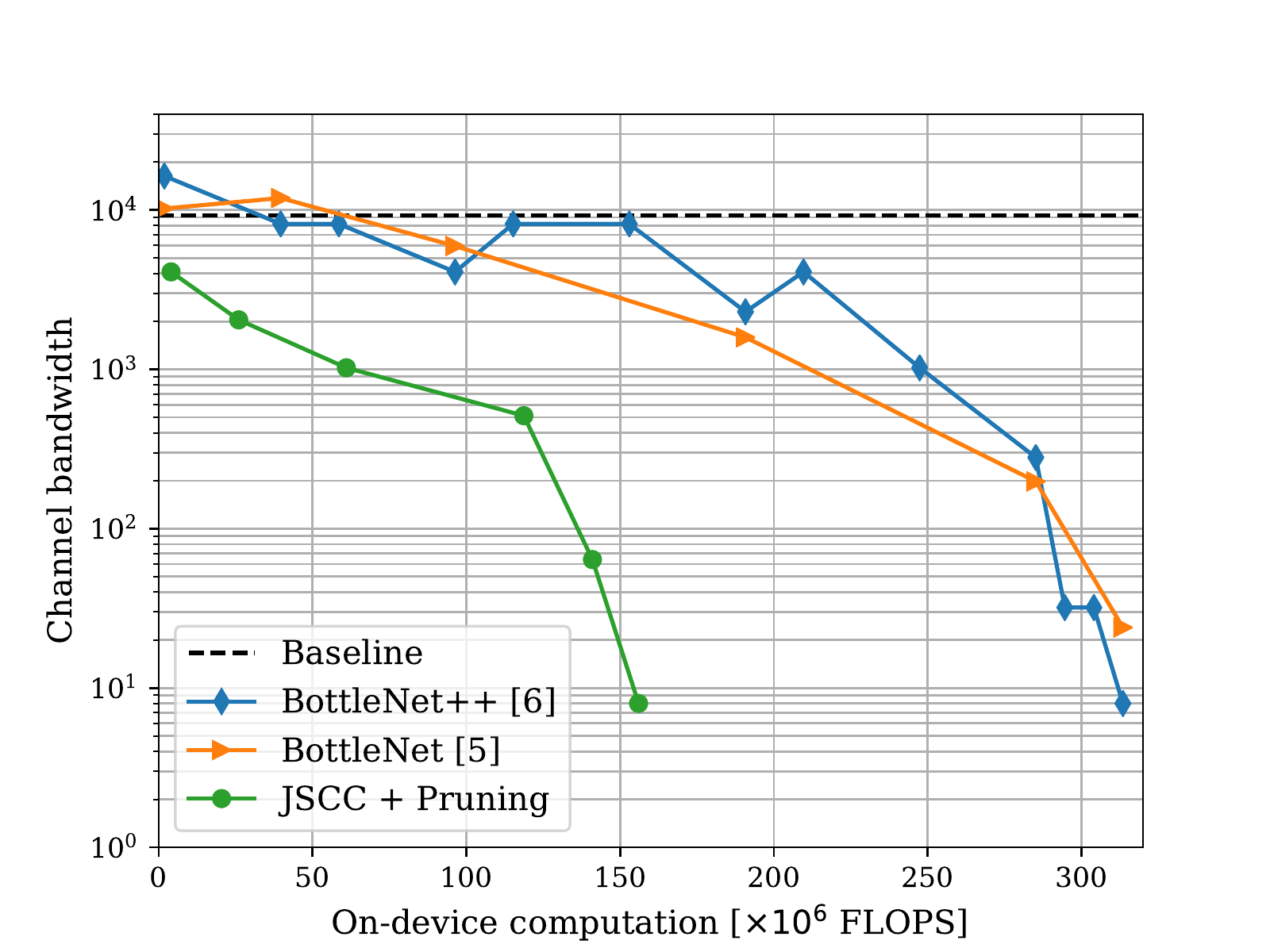}
\end{center}
\caption{Channel bandwidth as a function of on-device computation. The accuracy remains within $2\%$ from the classification baseline.}
\label{fig:comparison}
\end{figure}

In this section we select the models that minimize the channel bandwidth, which we define as the number of real symbols transmitted per image, and maximize the pruning ratio (which results in the minimal on-device computation), under a channel SNR of $14.5\mathrm{dB}$, allowing for a maximum drop of $2\%$ in the classification accuracy compared to the baseline. Results in Fig. \ref{fig:comparison} clearly show that our proposed approach beats both the JSCC-based BottleNet++ \cite{bn_plus_plus} and the digital communication based BottleNet \cite{bottlenet} schemes by a large margin. The proposed scheme requires only $4\times 10^6$ FLOPs to achieve approximately $3\times$ bandwidth reduction compared to the \textit{baseline}, which we define as transmitting the original PNG image and performing classification on the edge server without any processing at the edge device.

\begin{figure}[h]
    \centering
    \begin{subfigure}[t]{0.49\linewidth}
        \centering
        \includegraphics[width=\textwidth]{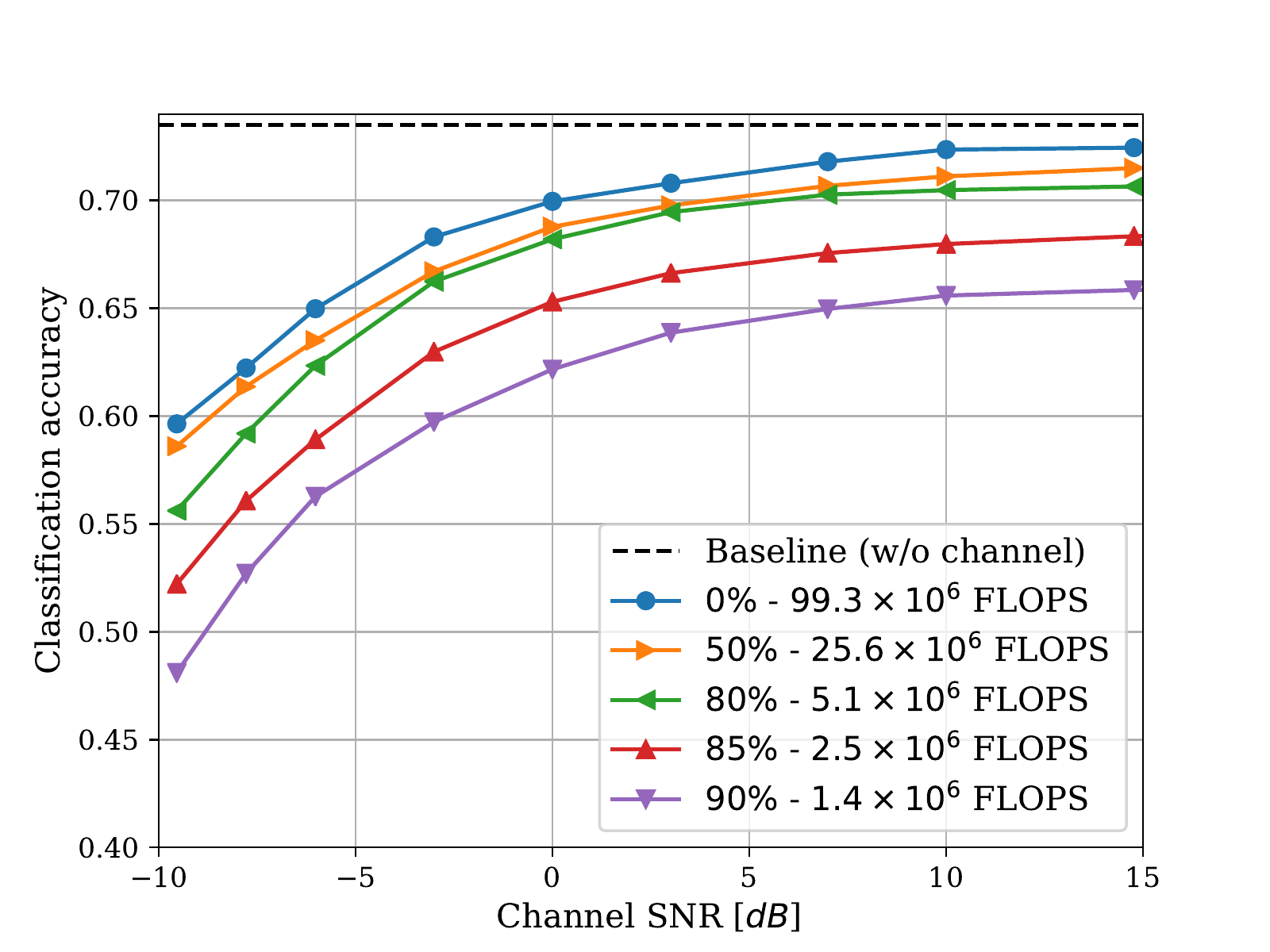}
        \caption{$B=2048$, splitting after pooling 2}
        \label{fig:pruning4}
    \end{subfigure}
    \begin{subfigure}[t]{0.49\linewidth}
        \centering
        \includegraphics[width=\textwidth]{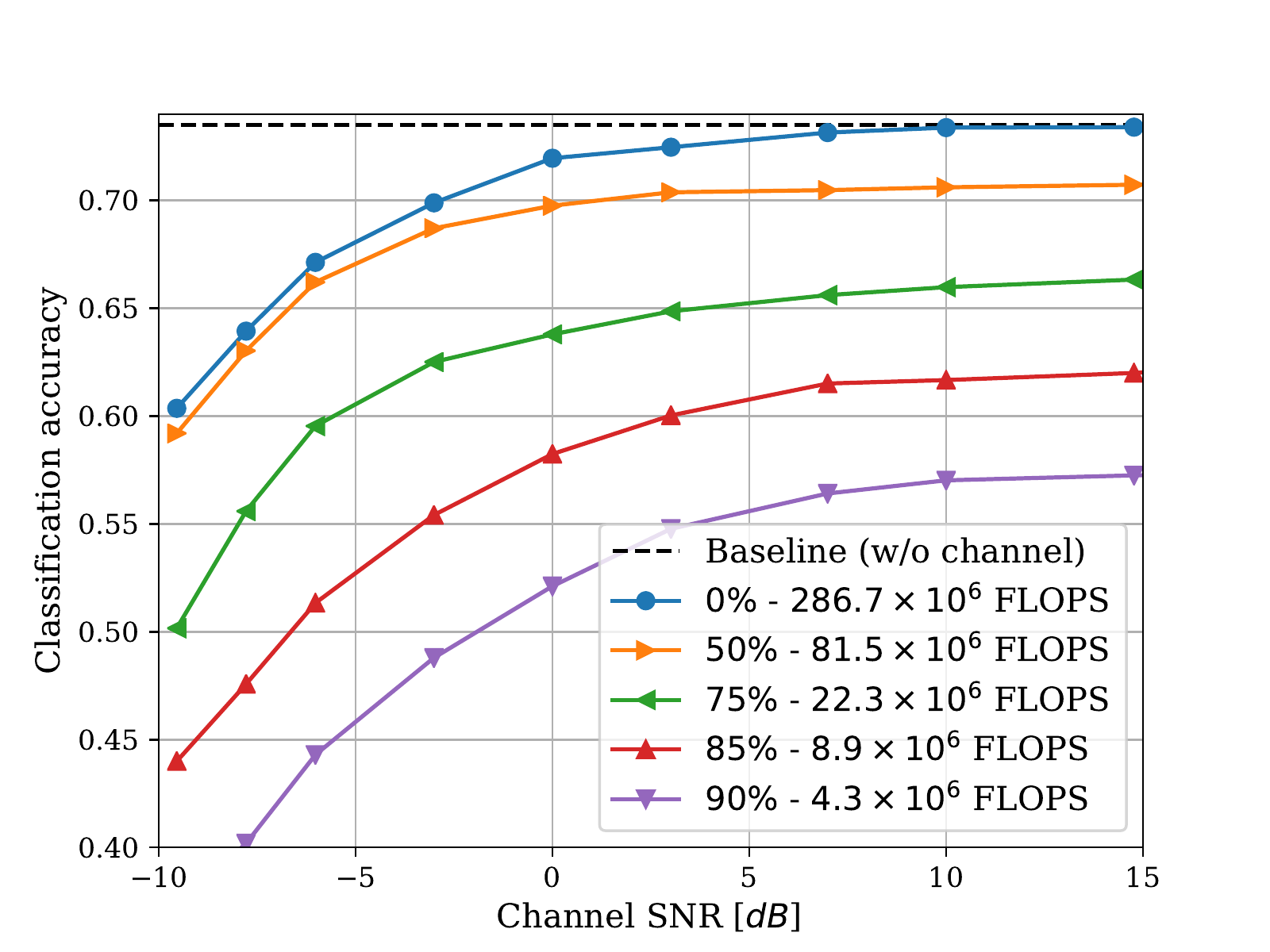}
        \caption{$B=128$, splitting after pooling 4}
        \label{fig:pruning10}
    \end{subfigure}
       
    \caption{Classification accuracy as a function of the channel SNR for different pruning ratios and channel bandwidth fixed to $B$.}
    \label{fig:pruning}
\end{figure}



Another superiority of our approach is that achieving $64\times$ bandwidth reduction (from 512 symbols to 8 symbols) after the last pooling in VGG16 network is possible with only $156\times 10^6$ FLOPs, which is only half of the operations necessary to run a single forward pass of the unpruned network. More importantly, given $156\times 10^6$ FLOPs limitation, which is the number of operations that can be performed on a single Apple Watch device within $0.05s$, we achieve $1024\times$ bandwidth reduction compared to \cite{bn_plus_plus}. 

\subsection{Comparison between different pruning ratios}

In this section we evaluate the influence of different pruning ratios on classification accuracy under different channel SNRs for fixed splitting points and available channel bandwidths (Fig. \ref{fig:pruning4} and Fig. \ref{fig:pruning10}).

It is clear, that pruning leads to a drop in accuracy, as expected. Nevertheless, given reasonable pruning ratios of up to $50\%$, accuracy drop decreases as we approach very low values of channel SNR. This behaviour may stem from the fact that feature distortion caused by network pruning becomes less significant when the channel is very noisy. Another important observation is that very high pruning ratios do not seem to reduce nor improve the robustness of the communication scheme - the accuracy drop caused by reducing the channel SNR follows a similar trend for every pruning ratio considered in this experiment.

\subsection{Comparison between different channel bandwidths}

\begin{figure}[h]
    \centering
    \begin{subfigure}[t]{0.49\linewidth}
        \centering
        \includegraphics[width=\textwidth]{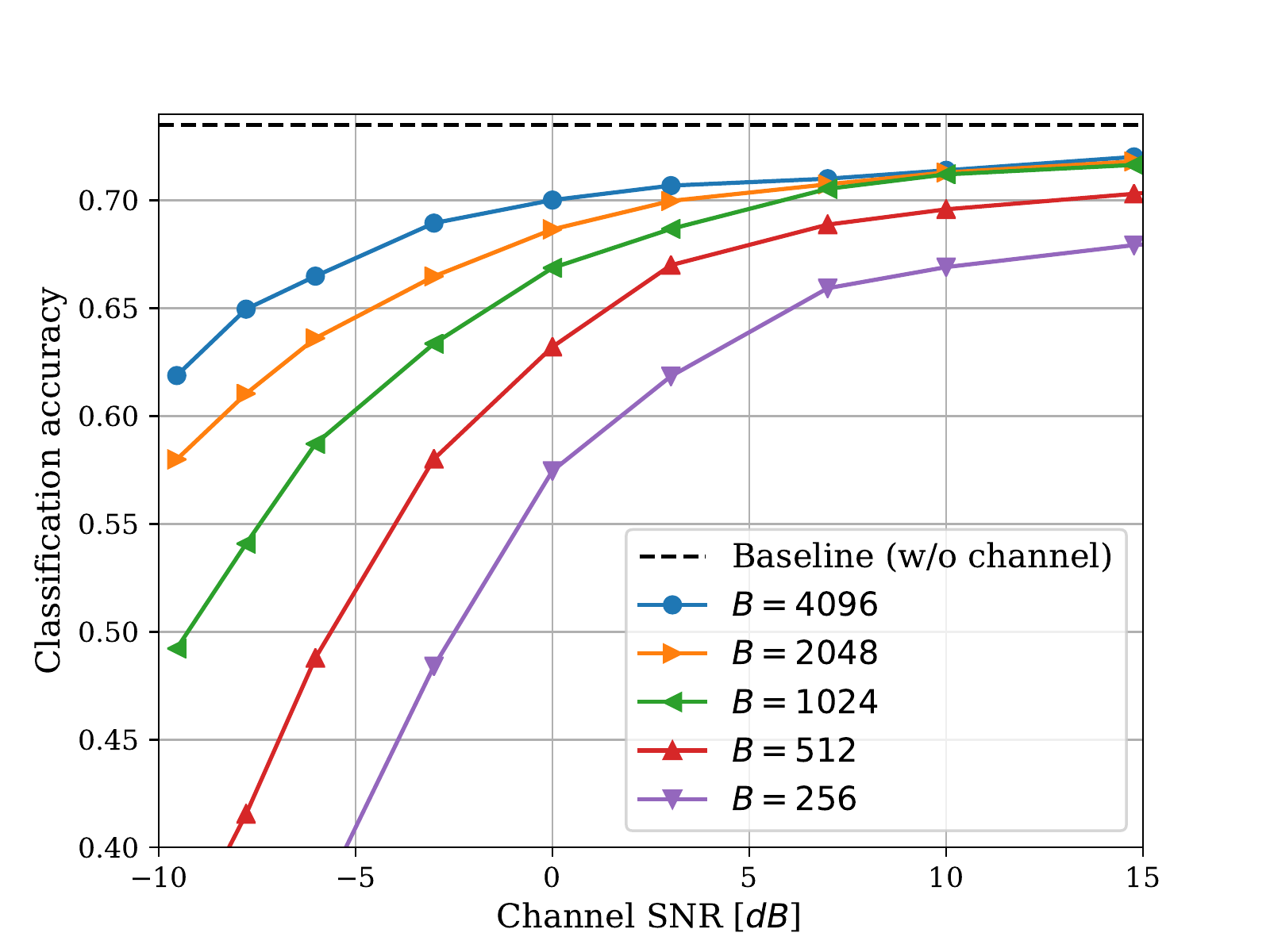}
        \caption{Splitting after pooling 2 with pruning ratio of $0.5$ ($24.3 \times 10^6$ FLOPs)}
        \label{fig:bandwidth4}
    \end{subfigure}
    \begin{subfigure}[t]{0.49\linewidth}
        \centering
        \includegraphics[width=\textwidth]{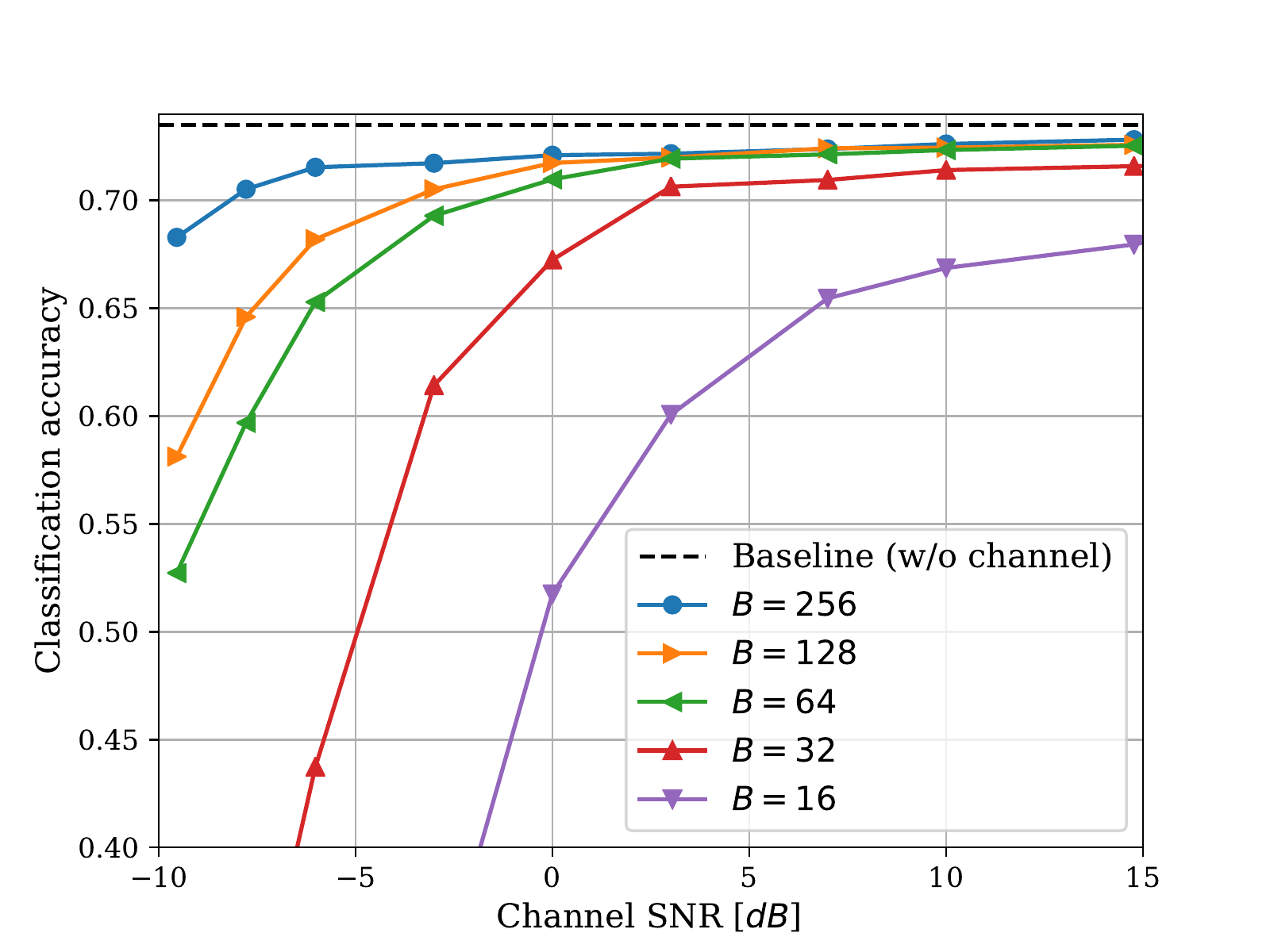}
        \caption{Splitting after pooling 4 with pruning ratio of $0.25$ ($160.2 \times 10^6$ FLOPs)}
    \label{fig:bandwidth10}
    \end{subfigure}
    \caption{Classification accuracy as a function of the channel SNR for different channel bandwidths.}
    \label{fig:bandwidth}
\end{figure}

In our last experiment, we fix the pruning ratio and the splitting point and examine the influence of the available bandwidth on the classification accuracy under different channel SNR values (Fig. \ref{fig:bandwidth}). One can clearly see that the available bandwidth is a crucial factor in the performance. In our experiments, reducing the bandwidth produced similar results for high SNR values. Nevertheless, the more limited the available bandwidth is, the sharper the drop in the accuracy with channel SNR.



\subsection{Selecting the optimal splitting point}

As demonstrated by the results, the selection of the optimal splitting point is a challenging and multi-dimensional problem. It is necessary to consider factors such as desired classification performance, the channel quality, available bandwidth, as well as the computational power of the edge device. For example, given the strict performance bounds in Fig. \ref{fig:comparison}, where each point of the \textit{JSCC + Pruning} curve corresponds to pruning layers up to pooling 1, 2, 3, 3 (higher pruning rate), 4, and 5, respectively, it is shown that changing the splitting point towards the end of the network decreases the bandwidth needed to send the information, but increases the required on-device computations, as more layers have to be processed by the edge device. Moreover, lowering the accuracy of the classification task allows to further reduce the computational requirements, or gain more robustness against channel noise.

\section{Conclusions}

We studied joint device-edge inference considering an IoT device with limited computational resources, and a wireless channel to the network edge. In particular, we considered image classification over a power and bandwidth limited edge device. We proposed pruning of the baseline taking into account the noise introduced over the channel, under a constraint on the available bandwidth. Our approach achieves superior results in classification accuracy even with extremely limited computational and communication resources.

\bibliographystyle{IEEEtran}
\bibliography{IEEEabrv,refs}

\end{document}